\documentclass[twocolumn,preprintnumbers,pre]{revtex4}
\usepackage{dcolumn}\usepackage{bm}
\usepackage{epsfig}
\usepackage{graphicx}

\begin{document}

\preprint{APS/PRE}
\title{Brownian motion of ellipsoidal particles on a granular magnetic bath}

\author{C. Tapia-Ignacio$^1$, R. E. Moctezuma$^2$, F. Donado$^1$,
and Eric R. Weeks$^{3}$}
\email{erweeks@emory.edu}
\affiliation{$^1$Instituto de Ciencias B\'asicas e Ingenier\'{\i}a, Universidad \\ Aut\'onoma del Estado de Hidalgo, Mineral de la Reforma 42184, Hidalgo, M\'exico, \\$^2$CONACYT-Instituto de F\'isica ``Manuel Sandoval Vallarta'', Universidad Aut\'onoma de San Luis Potos\'i, Alvaro Obreg\'on 64,
78000 San Luis Potos\'i, S.L.P., M\'exico\\
$^3$Physics Department, Emory University, Atlanta, Georgia 30322 USA}


\begin{abstract}
We study the Brownian motion of ellipsoidal particles lying on
an agitated granular bath composed of magnetic particles. 
We quantify the mobility of different floating ellipsoidal particles
using the mean square displacement and the mean square angular
displacement, and relate the diffusion coefficients to the bath
particle motion.  In terms of the particle major radius $R$, we find the translational diffusion coefficient scales roughly as $1/R^2$ and the rotational diffusion coefficient scales as roughly $1/R^4$; this is consistent with the assumption that diffusion arises from random kicks of the bath particles underneath the floating particle.  By varying the magnetic forcing, the bath particles diffusivity changes by a factor of ten; over this range, the translational and rotational diffusion of the floating particles change by a factor of fifty.  However, the ratio of the two diffusion constants for the floating particles is forcing-independent.   Unusual aspects of the floating particle motion
include non-Gaussian statistics for their displacements.
\end{abstract}

\pacs{}
\maketitle

\section{Introduction}

Brownian motion (diffusion) occurs in a wide variety of processes
in nature.  While initially observed in regular liquids, its theory
has been applied not only in physics, but in biology, chemistry, and
even economics.  In macroscopic granular systems, external forcing
can effectively ``thermalize'' the system and result in diffusive
motion \cite{feitosa02,ojha04,melby05,puckett13,fragility18,olafsen03,olafsen07}.
In super-cooled liquids, diffusion slows greatly,
but with rotational diffusion of non-spherical
probes slowing even more than translational diffusion
\cite{fujara1992,lohfink1992,biroli13,weeks12,weeks17}.  Some
granular experiments \cite{fragility18,marty05,keys07,xia15} and
simulations \cite{avila14,avila16} showed that granular systems
can exhibit glassy behavior under some conditions, with diffusive
motion greatly slowed.  However, rotational diffusion has not been
studied in these granular experiments.

In this paper, we study the translational and rotational
behavior of floating ellipsoidal particles on a magnetic
granular bath, and examine the influence of the bath particle
intensity on the motion of the floating particles.  By varying
the forcing applied to the magnetic bath particles, we can vary
the diffusivity of the floating particles by nearly two orders of
magnitude.  For low magnetic forcing, the bath particles exhibit  subdiffusive behavior somewhat like a supercooled fluid.  Our overall goal is to understand the behavior of the floating particles as subjected to this particular form of effective ``thermal'' forcing as we vary the motion of the bath particles; essentially, to add our previously described system \cite{fragility18,scientific,cecilio16} to the above mentioned examples of granular ``thermal baths''  \cite{feitosa02,ojha04,melby05,puckett13}.  Across the range of bath particle
conditions, we find that the floating particles always have
rotational diffusion and translational diffusion coupled, that is,
their ratio is constant.  While the coupling of the diffusion
constants is to be expected for particles in a ``normal'' fluid,
we find other unusual aspects of the floating particle motion such
as non-Gaussian statistics.  Evidence from examining different sizes of floating particles suggests that the diffusive motion can be understood as due to temporally periodic, randomly oriented kicks from the bath particles.

\section{Experimental setup}
\label{setupsection}

\begin{figure}[!t]
\centering
\psfig{file=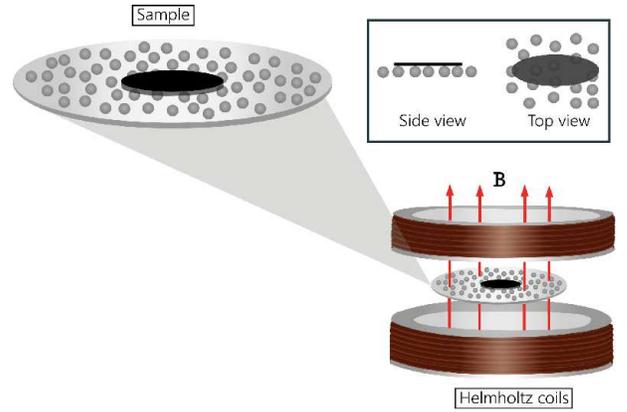,width=8.0cm} 
\caption{The experimental setup consists of magnetic particles
in a horizontal container (diameter 132 mm) in the middle of
Helmholtz coils (inner diameter 152 mm, outer diameter 179 mm). A sinusoidal signal fed into a power amplifier produces an alternating magnetic field. The spherical bath particles have permanent magnetic dipoles so the changing field causes the particles to roll on the surface in random directions.  A flat particle ``floats" on the bath particles and moves due to the random kicks it receives from the underlying bath particles.}
\label{setup}
\end{figure}

The experimental configuration consists of a flat particle
floating on a forced granular medium; see Fig.~\ref{setup}. This
granular medium is composed of a 2D ensemble of stainless steel
spherical particles of diameter 1 mm, mass 4.2 mg, and area
fraction 0.23.  The system is placed in a
time-dependent magnetic field.  The steel particles have been
previously magnetized so that they possess a permanent magnetic
dipole.  The magnetic field is generated by a pair of Helmholtz
coils fed by a power amplifier driven such that the magnetic field is $B_c + B_o \sin 2 \pi f t$, oriented vertically.  The oscillatory component causes the particles to
move randomly:  each steel particle tries to align its magnetic
dipole moment with the instantaneous field (pointing up or down),
causing the particle to rotate in a random direction to implement
the reorientation.  When the field reverses, the particle tries to
reorient, subject to the rotational inertia it already possesses.
The result is that the steel particle motion is quite random and
plays the role of a thermal bath.  The bath particles roll and slide, but always remain in contact with the surface; bouncing is not observed.

Decreasing $B_o$ toward $B_c$ results in an effectively ``cooler'' granular bath, as the particles spend less time during the cycle needing to reorient their dipoles.  The constant magnetic field component $B_c$ ensures that the particles spend more time with vertically oriented magnetic dipoles, and these aligned dipoles repel each other, thus frustrating the formation of permanent magnetic clusters which are observed in the absence of a constant magnetic field component \cite{voronoi18,donado17}.
Although this system is highly dissipative, it reaches a stationary
state which fits the necessary conditions to be considered as
an Ornstein-Uhlenbeck stochastic process \cite{scientific}. The
dynamics of a system similar to these bath particles has already
been studied, and it has been demonstrated that for certain
conditions the magnetic field is proportional to an effective
temperature \cite{cecilio16}.

For all experiments, we use $B_c = 33$~G for the constant component.  The intensity of the bath particle
motion is varied with the magnitude of the oscillating magnetic
field: $B_{o}=$ 66, 60, 54, 48, 42, 36 and 30 G,  all of them with
a frequency of $f=10$~Hz. In this system, for high intensities of the
field, the bath particles show the dynamics similar to that of a
Brownian gas, and the velocity distributions are Maxwell-Boltzmann
distributions \cite{cecilio16}. For low intensities, the motion of
the particles decreases, showing freezing-like effects resembling
the dynamics near the glass transition \cite{fragility18}.  For the lowest amplitude we consider ($B_o=30$~G), we have $B_o < B_c$ and thus the imposed magnetic field never changes direction; the magnetic dipoles do not need to rotate.  Nonetheless, the particles are observed to still rotate and move, albeit slowly.
This may be due to two possibilities.  First, the Earth's magnetic field (measured to be 0.6 G) is at an angle to the imposed vertical field, and that slight off-vertical component to the field perhaps causes some particle rotation at the times when the absolute magnitude of the imposed magnetic field is the smallest.  Second, the oscillating magnetic field likely gives a time varying component to the particles' magnetic dipoles, causing a time varying repulsive interaction and thus causing the particles to move.  Evidence for this second mechanism is that at lower area fractions, the particles are no longer observed to move other than small vibrations.  At the area fraction used in this work ($\phi = 0.23$),
when $B_{o}$ is less than 30~G, the spatial distribution of the bath particles becomes inhomogeneous during the course of the experiment, thus preventing us from conducting long duration experiments at low bath particle motion.

We conduct two series of experiments.  One is a study of the size dependence of the floating particle motion, and the other is a study of the forcing dependence of the floating particle motion.  These two series of experiments use different bath particles.  By chance, the bath particles for the size dependence experiments are less magnetized than the bath particles for the forcing dependence.  For this reason, while we report the amplitude of the oscillatory magnetic field $B_o$ for the results discussed in the next section, a more useful characterization is the length scale $L$ defined by the bath particle mean square displacement:  $L^2 = \langle \Delta r^2 \rangle (\Delta t = 0.1$~s).  The time scale $\Delta t = 0.1$~s is chosen to match the forcing frequency; further justification for this choice is presented below.

\begin{figure}[!t]
\centering
\psfig{file=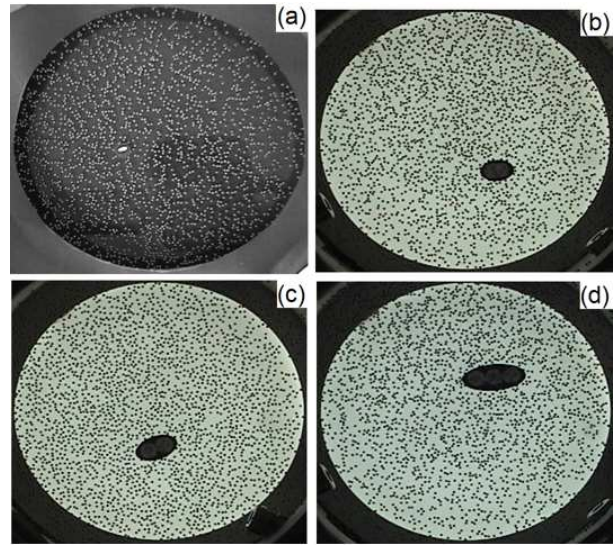,width=8.0cm}
\caption{Images showing floating particles with aspect ratio (a)
$\alpha = 1.00$, (b) $\alpha = 1.44$, (c) $\alpha = 1.48$, and
(d) $\alpha = 2.20$.  The bath particles are the smaller spheres.
Note that the image in (a) is viewed with reflected light, while
the other are viewed with transmitted light, leading to the
different appearance of (a) from the other panels.  
The circular particle in (a) has a white ellipse drawn on it to facilitate rotational tracking of this otherwise circularly symmetric particle.}
\label{pictures}
\end{figure}

We use floating particles with aspect ratios ranging from $\alpha = 1.00$ to $\alpha = 2.20$, as shown in  Fig.~\ref{pictures}. Further
details of these particles are listed in Table~\ref{tracers}.
The floating particles are non-magnetic, and do not interact
with the magnetic field.  Rather, they sit atop the magnetic bath
particles, and move due to the random kicks of the bath particles.  The floating particles experience frictional forces from the bath particles, and also the bath particles occasionally collide with the edges of the floating particles.  The floating particles are sufficiently stiff that they remain flat, and essentially horizontal.  In particular, we avoid using smaller floating particles which would tilt out of plane.  Observations of the bath particles underneath the floating particles are possible, as the floating particles are semi-transparent; we find the bath particles underneath have a mean square displacement roughly 80\% the magnitude of the particles away from the floating particle.  We observe that particles enter and exit the floating particle region constantly throughout the experiments.

For the size dependence experiments, we obtain one three-minute video for each floating particle.  For the forcing dependence experiments, we obtain three series of three-minute videos for each floating
particle and magnetic forcing amplitude.  From each video we track
the translational and rotational motion of the floating particle,
along with the bath particle motion \cite{mosaic}.  Our temporal
resolution is 1/60~s, our spatial resolution is better than
0.07~mm ($x$ and $y$ position), and our angular resolution is
better than 0.006~rad ($\approx 0.4^\circ$).  As is apparent from Fig.~\ref{pictures}, the images are slightly compressed in the vertical direction, which is accounted for in the analysis (both in the translational calibration and the angular calibration).  Even at the highest level of forcing, it is rare for the floating particle to move close to the container boundary within three minutes.  We exclude movies where this occurs from the data analysis.

\begin{table}[t]
\begin{center}
\begin{tabular}{ccccc}
aspect ratio & $2r_{\rm major}$ & $2r_{\rm minor}$ & mass \\
\hline
1.00 & 6.4 mm & 6.4 mm & 2.2 mg \\
1.44 & 13.0 mm & 9.0 mm & 7.1 mg \\
1.48 & 17.0 mm & 11.5 mm & 10.4 mg \\
2.20 & 24.4 mm & 11.1 mm & 17.0 mg \\
\end{tabular}
\end{center}
\caption{
Details of the floating particles, including semi-major diameter
and semi-minor diameter.  The particles are cut from paper with a thickness of 0.12~mm and an area density of 0.073 mg/mm$^2$.
}
\label{tracers}
\end{table}

\section{Results}

\subsection{Size dependence}

\begin{figure}[!b]
\centering
\psfig{file=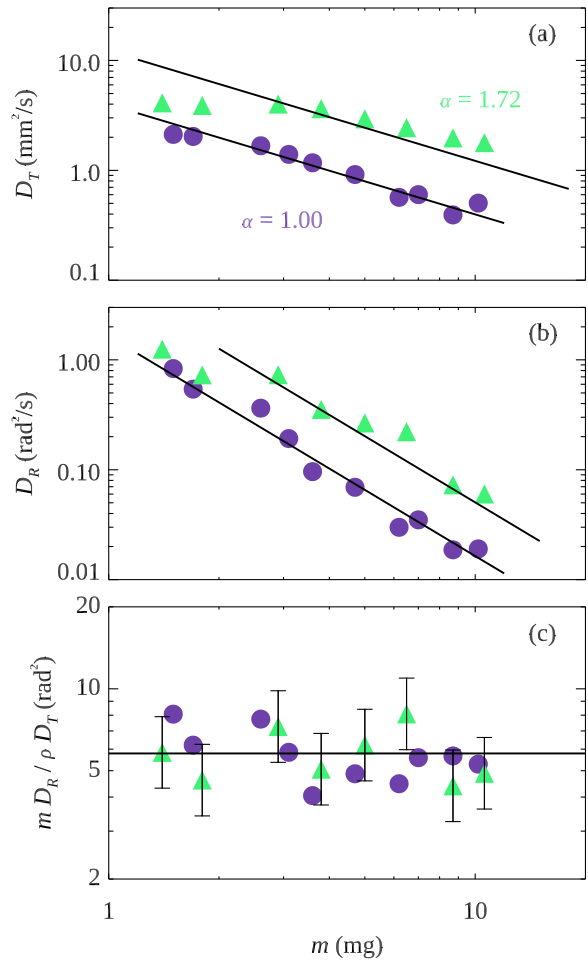,width=8.0cm}
\caption{(a) Translational diffusion coefficient $D_T$ as a function of floating particle mass for ellipses (aspect ratio $\alpha=1.72$) and circular particles ($\alpha=1.00$) as labeled. The fit lines are constrained to follow $D_T \sim 1/m$.  Unconstrained least squares fits give power law exponents of -0.9 (circular particles) and -0.4 (ellipses).  (b) Rotational diffusion constants $D_R$ for the same particles.  The fit lines are constrained to follow $D \sim 1/m^2$.  Unconstrained least squares fits give power law exponents of -2.1 (circular particles) and -1.5 (ellipses).  The uncertainties of the diffusivity data in panels (a) and (b) are the symbol size.  (c) Ratio $m D_R / \rho D_T$ as a function of mass $m$.  The horizontal line shows the mean value for all data, $5.8 \pm 1.3$~rad$^2$ (mean $\pm$ standard deviation).  The uncertainty of the ratio data is $\pm 35$\%, with representative error bars for the ellipse data.
}
\label{size}
\end{figure}

We first wish to understand the dependence of diffusion on the floating particle size.  To do this, we 
use our largest forcing ($B_o = 66$~G) although with the less magnetized bath particles.  The motion of the bath particles over a 0.1~s time scale is $L = \sqrt{\langle \Delta r^2 \rangle} = 1.6$~mm.  In this situation, the bath particles are diffusive at time scales $\Delta t > 1$~s.
We find that the floating particles are also diffusive, and so we study how the diffusion coefficient depends on particle size.  
In a 3D Newtonian fluid, the rotational motion of ellipsoids follows $D_T \sim 1/r_{\rm major}, D_R \sim 1/r_{\rm major}^3$, in part because the two diffusion coefficients have different units \cite{sutherland05,einstein05,debye29}.  In our experiments it is more precise to measure the floating particles' masses rather than their radii, so accordingly we plot $D_T$ and $D_R$ as functions of mass $m$ in Fig.~\ref{size}.  We find roughly that $D_T \sim m^{-1}$ and $D_R \sim m^{-2}$, as shown by the superimposed lines in Fig.~\ref{size}.  As $m \sim$ area $\sim r^2$, this shows that the diffusivity behaves differently in our experiment than the 3D Brownian motion case, although the ratio $D_T/D_R$ scales as $\sim r^2$ for our experiment as is also the case for 3D Brownian diffusion.  Indeed, the ratio $m D_T/\rho D_R$ is plotted in Fig.~\ref{size}(c) and within uncertainty appears independent of $m$; this uses the area density of the paper ($\rho = 0.073$~mg/mm$^2$) to simplify the units.  Within the uncertainty, this ratio is independent of the aspect ratio [mean ratio is $5.78 \pm 1.29$ for $\alpha=1.00$, $5.80 \pm 1.32$ for $\alpha=1.72$, (mean plus or minus standard deviation)].

The $1/r^2$ dependence of the translational diffusion coefficient is plausibly due to the random kicks of the tracer particles.  As noted above, the bath particles have an area fraction $\phi=0.23$.  An elliptical floating particle with aspect ratio $\alpha$ and minor axis radius $r$ will cover $N = \alpha \pi r^2 \phi$ bath particles on average.  During one period of the oscillatory magnetic field ($\tau=0.1$~s), each bath particle kicks the floating particle in a random direction.  Adding these kicks randomly, the net force should be proportional to $(\alpha \pi r^2 \phi)^{1/2} \sim r$.  The mass of the particle is proportional to the area which is proportional to $r^2$, so the acceleration of the floating particle during $\tau$ is proportional to $1/r$.  The displacement of the floating particle during that time is proportional to acceleration and thus $1/r$.  Assuming that the bath particle motion is random each period, the Central Limit Theorem yields a diffusion coefficient $D_T =$ (displacement)$^2/2\tau \sim 1/r^2 \sim 1/m$, in agreement with the results of Fig.~\ref{size}(a).  The data show this argument is a bit simplistic, as it would predict that $D_T$ does not depend on aspect ratio, whereas Fig.~\ref{size}(a) shows that for the same mass, the elliptical particle has a larger translational diffusion constant (by a factor of 3.1).

A similar line of reasoning for the rotational motion leads to $D_R \sim 1/r^4$.  A bath particle exerts a torque proportional to $r$, and assuming again the $N$ torques add randomly, the net torque should be proportional to $r^2$.  The moment of inertia for a circular disk is $I=mr^2/2 \sim r^4$, so the angular acceleration and thus angular displacement during one oscillation period scale as $1/r^2$.  This leads to $D_R \sim 1/r^4 \sim 1/m^2$, the relationship indicated by the solid lines in Fig.~\ref{size}(b).  This works very well for the circular floating particles (circle symbols in the figure), and less well for the elliptical floating particles (triangle symbols).  The deviations suggest that the interaction between the bath particles and the floating particles may be more complex than we assume here.  For $D_R$, we find the ellipses diffuse a factor of 3.1 times faster than the circles for equivalent $m$, the same factor as for $D_T$.

\subsection{Varying forcing intensity}

\begin{figure}[!t]
\centering
\psfig{file=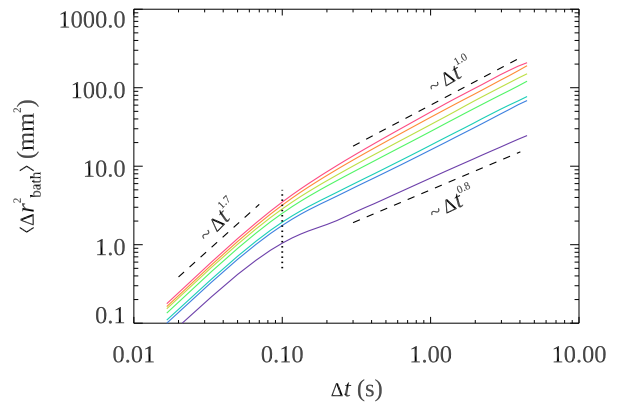,width=8.0cm}
\caption{Mean square displacement $\langle\Delta r^2_{\rm bath}
\rangle$ for the bath particles.  The vertical dotted line
indicates the time scale (0.1 s) corresponding to the period of
the magnetic forcing.  The intersection of the data with this
vertical line sets the length scale $L$ through $L^2 = \langle\Delta
r^2\rangle (\Delta t = 0.1 {\rm s})$. The dashed lines indicate power
law growth with the exponents as labeled.  The lowest forcing (30 G) of the bath
particles corresponds to the purple curve at the bottom 
and increases up to 66 G (red curve at the top).
}
\label{msd_bath}
\end{figure}

We now turn to the experiments varying the forcing intensity.  As noted in Sec.~\ref{setupsection}, these experiments are done with bath particles that are more strongly magnetized and which thus can be agitated quite strongly by the oscillatory magnetic field.
To characterize the motion of the bath particles, we calculate
the mean square displacement (MSD) for each magnetic field amplitude
we consider.  The results are shown in Fig.~\ref{msd_bath}.  It is
interesting to note that for all of the data, at short $\Delta
t$, the MSD of the bath particles is superdiffusive with $\langle
\Delta r^2 \rangle \sim \Delta
t^{1.7}$.  This implies that over a short time scale particles
are more likely to move in straight lines, rather than being
purely diffusive random walks \cite{bouchaud90}.
At high magnetic field
intensities, the motion of the bath particles is diffusive (top
curves of Fig.~\ref{msd_bath}).  As the magnetic field intensity
decreases, bath particle motion is slower.  At the lowest forcing,
the MSD has a short subdiffusive bend around $0.1 < \Delta t < 0.3$~s (bottom curve of
Fig.~\ref{msd_bath}).  Even at longer time scales ($\Delta t > 1$~s) it does not quite achieve diffusive behavior, but rather follows $\langle \Delta r^2 \rangle \sim \Delta t^{0.8}$.  This resembles a supercooled fluid close
to the glass transition
\cite{kob95a,hunter12,reis07,magdaleno13}.

\begin{figure}[!t]
\centering
\psfig{file=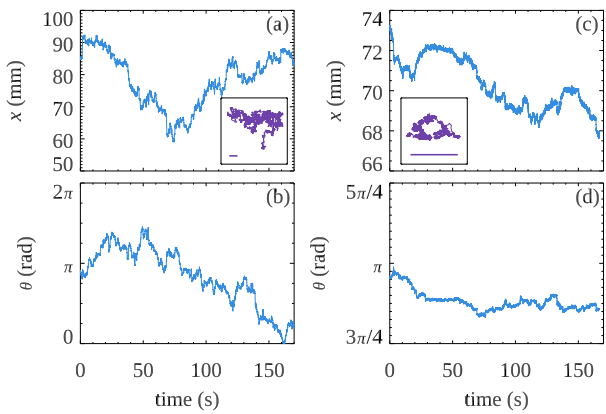,width=8.0cm}
\caption{Trajectories for the floating particle with aspect ratio $\alpha =2.20$ for a high excitation case with $L = 1.68$ mm (a,b) and a low excitation case with $L = 0.92$ mm (c,d).  Panels (a,c) show the $x$ position as a function of time.  The insets show the full trajectory in $x$ and $y$ with scale bars of length 5~mm.  Panels (b,d) show the corresponding angular orientation as a function of time.  Note the vertical scales are significantly reduced in panels (c,d) as compared to (a,b).}
\label{traj}
\end{figure}

We likewise consider the floating particle motion, which has both
translational and rotational components.  Example translational and
rotational trajectories of a floating particle with aspect ratio
$\alpha=2.20$ for a high and a low excitation case are shown in
Fig.~\ref{traj}. Both the rotational orientation and the position
in $x$ show a greater variation for higher excitations than for
the lower ones. The full trajectories for high excitations expand
noticeably over a larger area compared with the trajectories at
low excitations (insets of Fig.~\ref{traj}).

\begin{figure}[!t]
\centering
\psfig{file=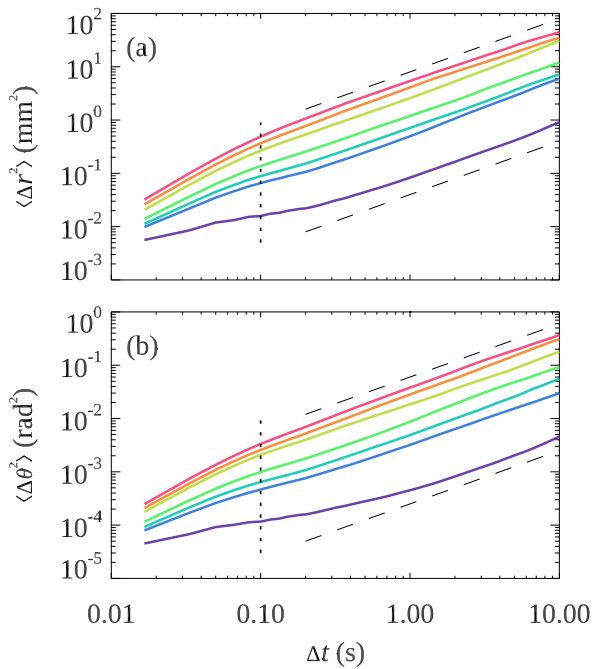,width=8.0cm}
\caption{(a) Mean square displacement $\langle\Delta r^2\rangle$
for the longest floating particle $(\alpha = 2.20)$.  The color
of each curve relates to the intensity of the bath particle
motion; see text for details.  The vertical dotted line indicates
the time scale ($0.1$~s) corresponding to the period of the
magnetic forcing. The dashed lines indicate linear dependence
$\langle\Delta r^2\rangle \sim\Delta t$.  (b) Mean square angular
displacement $\langle\Delta\theta^2 \rangle$.  The colors are the
same as in (a).}
\label{msd_tracer}
\end{figure}

Figure \ref{msd_tracer} shows the MSD $\langle\Delta r^2\rangle$
and the mean square angular displacement (MSAD) 
$\langle\Delta\theta^2\rangle$ for the floating particle
with aspect ratio $\alpha = 2.20$.   The vertical dotted line
corresponds to the period of the magnetic forcing, and coincides
with a crossover region from short time scale behavior to long
time scale behavior.  In all cases there is diffusive behavior at large time
scales for both MSD and MSAD, regardless of the excitation of
the bath. This result is interesting, since it indicates that,
although the bath particles goes from diffusive to subdiffusive
as we decrease the magnetic forcing (Fig.~\ref{msd_bath}), the
floating particles maintain their diffusive behavior.  Intriguingly,
the bath particle MSDs all have a constant initial power law growth
(Fig.~\ref{msd_bath}, constant initial slope on the log-log plot)
whereas the floating particle MSDs exhibit an initial slope
dependent on the forcing.  Thus, within the time scales $\Delta
t < 0.1$~s (the forcing period), the coupling between the bath
particles and floating particles is nontrivial.  The $\Delta t
\rightarrow \infty$ behavior is also interesting:  for the bath
particles, the log-log slope of the MSD changes with the magnetic forcing
intensity, whereas for the floating particles their behavior is
always diffusive at long time scales (slope $=1$ on the log-log
plot).

\begin{figure}[!t]
\centering
\psfig{file=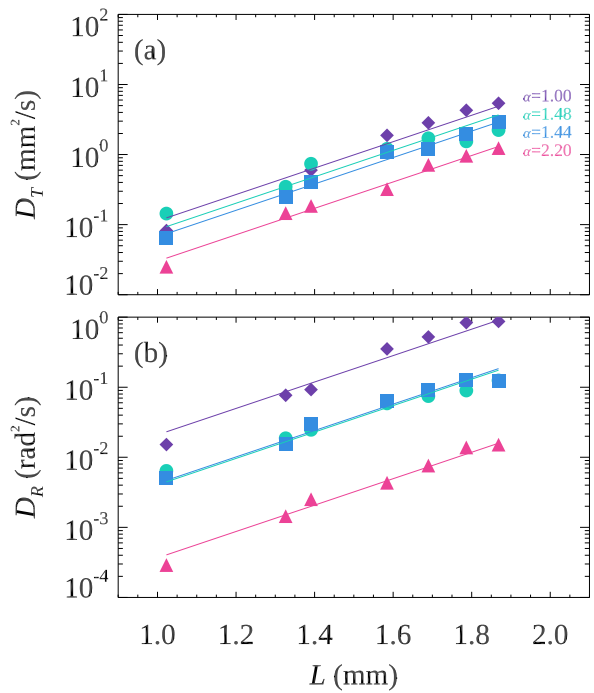,width=8.0cm}
\caption{(a) Translational diffusion coefficients $D_T$ as a
function of the forcing $L$.  The different symbols are for
different aspect ratio floating particles as indicated.  (b)
Rotational diffusion coefficients $D_R$ as a function of the
forcing $L$.  The symbols are the same as (a).  For both panels,
all fit lines have a constant slope:  $\ln(D) \sim L/L_1$ with
$L_1=0.23$~mm.  The uncertainty in $D$ is the symbol size.
} 
\label{diffusion}
\end{figure}

The differing behavior at long time scales between the bath
particles and the floating particles suggests that the direct
influence of the bath is only at shorter time scales; that the
diffusive behavior of the floating particles is emergent from the
random steps that occur for each cycle of the oscillatory magnetic forcing.  To investigate
this, we fit the $\Delta t \rightarrow \infty$ portion of the
floating particle MSD and MSAD curves to determine the diffusion
coefficients through $\langle \Delta r^2 \rangle = 4 D_T \Delta
t$, $\langle \Delta \theta^2 \rangle = 2 D_R \Delta t$.  We then
examine the dependence of the measured diffusion coefficients on
various quantities related to the bath particles.  In particular
we use the bath particle motions (quantified by the MSD) using
a natural choice for the time scale, $\Delta t = 0.1$~s (the period
of the magnetic forcing).  We define a length scale $L = \sqrt{
\langle \Delta r^2_{\rm bath} \rangle }$ using displacements over
this time scale.  Figure~\ref{diffusion} shows a linear relationship
between $L$ and the logarithms of all of the diffusion coefficients.
This relationship, $\ln(D) \sim L/L_1$, holds for all our results
with a common length scale $L_1 = 0.23$~mm; this is clear from
the least square fit lines in Fig.~\ref{diffusion} where the fit
is constrained to have a common slope and differing intercepts.
Defining $L$ with a different $\Delta t$ gives results that have
a larger least square fit error, reinforcing our choice to use $\Delta t = 0.1$~s.

We note that fitting the diffusion coefficients to $\ln(D) \sim
L^2 / L_2^2$ works nearly as well, with $L_2 = 0.82$~mm.  The
least square error is slightly larger, but the overall appearance
is similar to the results shown in Fig.~\ref{diffusion}.  The two
fitting constants $L_1$ and $L_2$ are both similar in order of
magnitude to the bath particle radius 0.5~mm.  The dependence of
$\ln(D)$ on $L$ or $L^2$ suggests something like an activated
process, although it is not apparent why this would be the case.


\begin{figure}[!t]
\centering
\psfig{file=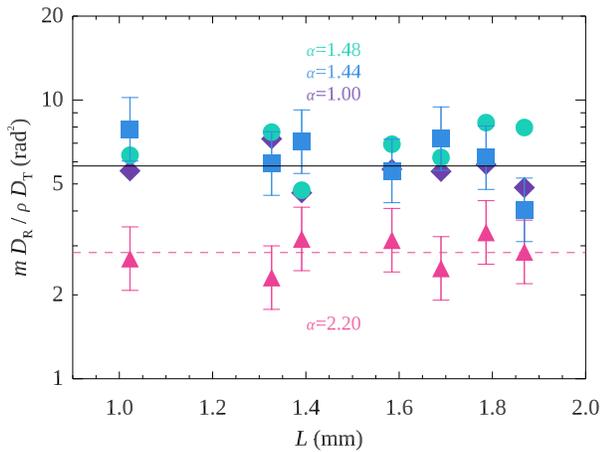,width=8.0cm}
\caption{The ratio of diffusion coefficients for floating particles with aspect
ratios as indicated, as a function of the forcing intensity $L$.
The ratio is nondimensionalized by multiplying by $m/\rho$, which is the area of the floating particle.
Representative error bars are shown for the $\alpha=1.44, 2.20$ data
sets; the
uncertainties of the other data sets are identical.  The solid horizontal line indicates 5.8~rad$^2$ [taken from Fig.~\ref{size}(c)] and the dashed horizontal line indicates 2.8~rad$^2$ (the mean value for the $\alpha=2.20$ data).  Taken individually, the mean values for $\alpha=1.00, 1.44, 1.48$ are $5.6, 6.2, 6.8$~rad$^2$.
}
\label{diffusion_rate}
\end{figure}

We next wish to compare $D_T$ to $D_R$.  As noted above, for diffusion in three dimensions, one expects $D_R / D_T \sim 1/r_{\rm
major}^2$.  For our particular particles Fig.~\ref{size}(c) shows that $m D_R / \rho D_T = A D_R / D_T$ collapses the data for different sizes of floating particles in terms of floating particle mass $m$, floating particle area density $\rho$, and area of floating particle $A$.  Accordingly, we
consider this non-dimensional ratio 
plotted as a function of $L$ in Fig.~\ref{diffusion_rate}.
Given that $D_T$ and $D_R$ have the same functional dependence on
$L$ (Fig.~\ref{diffusion}), it must be true that their ratio is a
constant, as Fig.~\ref{diffusion_rate} shows (within the
uncertainty).  For $\alpha=1.00, 1.44$, and $1.48$, the ratios agree quite well with the result from Fig.~\ref{size}(c), with $m D_R / \rho D_T\approx 6$.  For $\alpha=2.20$, this ratio is about half the value of the other experiments; it is unclear why this is the case.  To compare to the 3D result for spheres, we can use $A = \pi r^2$ to convert our ratio for the $\alpha=1.00$ particle, leading to a value of $r_{\rm major}^2 D_R/D_T = 1.8$.
For spheres diffusing
in 3D, this ratio would be 3/4
\cite{sutherland05,einstein05,debye29}, so a somewhat similar order of
magnitude.  More generally, we have $A \equiv \pi r_{\rm major}^2 / \alpha$, so the ratio $\beta = r_{\rm major}^2 D_R / D_T$ = 1.8, 1.4, 1.5, 0.4 for $\alpha = 1.00, 1.44, 1.48, 2.20$.  This decreasing trend with
increasing aspect ratio qualitatively disagrees with numerical results for 3D rod
diffusion derived by Tirado and Garcia de la Torre \cite{tirado79,tirado80}.  For example, they found that when the aspect ratio of a rod is increased from 2 to 4, $\beta_4/\beta_2 = 1.4$, an increase.  In contrast, when our aspect ratio is increased from 1.44 to 2.20, $\beta_{2.20}/\beta_{1.44} = 0.3$, a decrease.  This disagreement suggests,
not surprisingly, that the mechanism leading to diffusion in our
2D granular experiment are qualitatively different from normal 3D
diffusion in liquids.

According to the classic Stokes-Einstein-Sutherland and
Stokes-Einstein-Debye equations for $D_T$ and $D_R$ respectively,
the ratio $D_T/D_R$ should be a constant independent of temperature
and viscosity; the exact constant will depend on the floating
particle shape.  In glass-forming systems, prior observations
found ``decoupling'' of translational and rotational diffusion,
such that $D_R/D_T$ is not constant but rather decreases as the
glass transition is approached.  This has been seen in experiments
with small molecule glass-formers \cite{stillinger94,chang1994,cicerone1996}, colloidal glass experiments
\cite{weeks12,weeks17,pavlik02}, and simulations \cite{mazza06,chong05,chong09,kumar06}.  Our bath particles appear a bit like they are approaching a glass transition as the magnetic forcing is decreased, as seen by the subdiffusive behavior in
Fig.~\ref{msd_bath} for the lowest forcing.  Despite our bath particles becoming subdiffusive,
in our experiments we do not observe decoupling of the translational
and rotational diffusion:  there is no systematic dependence of
the data in Fig.~\ref{diffusion_rate}.  This is further evidence
that the long-time behavior of the bath particles does not
connect with the long-time behavior of the floating particles.

\begin{figure}[!t]
\centering
\psfig{file=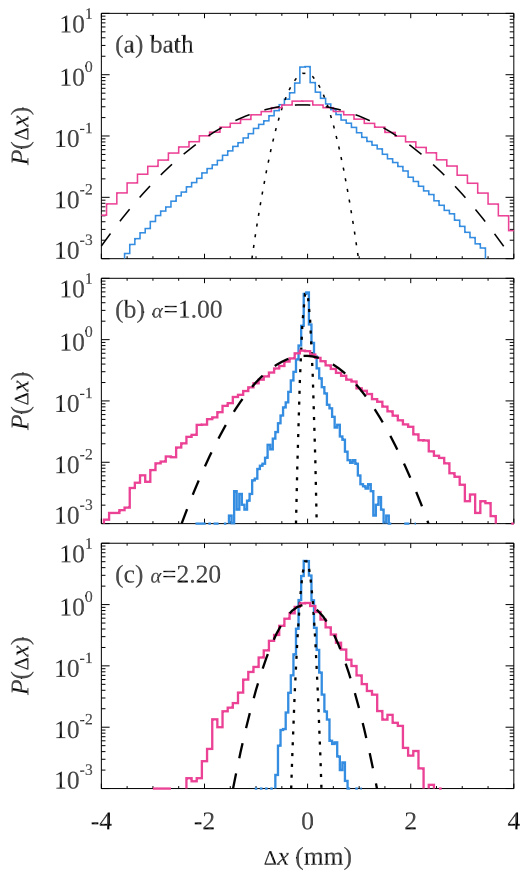,width=7.0cm}
\caption{Probability distribution functions $P(\Delta x)$ for (a)
bath particles, (b) floating particles with aspect ratio $\alpha$
= 1.00, and (c) floating particles with aspect ratio $\alpha$
= 2.20.  These are laboratory frame displacements, similar to the data shown in Fig.~\ref{traj}.
In each case the blue data correspond to $L = 0.92$~mm (the
lowest excitation) and the red data correspond to $L = 1.68$~mm (the
highest excitation).  $\Delta t = 0.1$~s is used to define the displacements.
The dashed and dotted lines are Gaussian fits
with the same standard deviation as the experimental data.
The non-Gaussian parameter
$\alpha_2$ has the values (a) 1.1, 0.1; (b) 6.1, 0.5; (c) 2.3,
0.6 for small and large $L$.}
\label{distributions}
\end{figure}

To better understand particle motion (both bath and floating
particles), we 
show representative probability distribution
functions of displacements in Fig.~\ref{distributions}.  In each
panel the blue data are for the lowest excitation and are the narrow
distributions; the red data are for the highest excitation and are
the broad distributions.  The dotted and dashed black lines are
Gaussian fits, and it is immediately apparent that the distributions
are broader than a Gaussian.  That is, large displacements are rare,
but more common than would be expected if the distribution was a
Gaussian of the same standard deviation.  Non-Gaussian distributions
are often seen in supercooled systems
\cite{olafsen99,kob95a,hunter12}, so the result is plausible
for the bath particles.  It is intriguing
that the distribution functions are non-Gaussian for the floating
particles even though these particles are diffusive at long time
scales (Fig.~\ref{msd_tracer}).  That is, often strong non-Gaussian
behavior is associated with a more pronounced plateau in the mean
square displacements; Fig.~\ref{msd_tracer} shows a weak plateau,
which is most obvious for the lowest forcing data.  In all cases,
the distributions are close to Gaussian for higher excitation.
The same qualitative results are found for the rotational
displacements (data not shown).


The relative width of the displacement probability distributions can
be quantified with the non-Gaussian parameter 
$\alpha_2(\Delta t)$ \cite{rahman64},
\begin{equation}
\alpha_2(\Delta t)=\frac{\langle\Delta r^4\rangle(\Delta t)}
{2[\langle\Delta r^2\rangle (\Delta t)]^2}-1 .
\end{equation}
This quantity is zero for a Gaussian distribution, and positive
for distributions with broad tails such as those shown in
Fig.~\ref{distributions}.  The specific values of $\alpha_2$ are
given in the caption to the figure.

\begin{figure}[!t]
\centering
\psfig{file=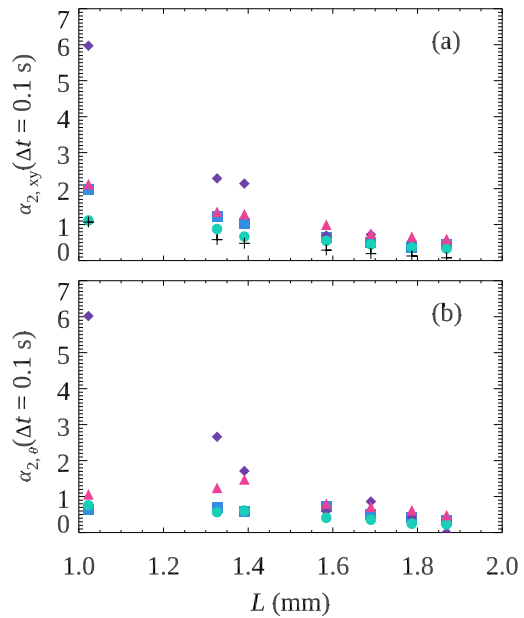,width=7.0cm}
\caption{(a) Non-Gaussian parameter for translational motion, evaluated at time scale $\Delta$t = 0.1 s.  The symbols have the same meaning as in prior figures.  The + symbols are for the bath particles.  (b) The same, for rotational motion.}
\label{non_gaussian}
\end{figure}

The non-Gaussian parameter is shown in Fig.~\ref{non_gaussian}(a)
for translational motion, and (b) for rotational motion, as a
function of the forcing strength $L$.  For these results, the time
scale $\Delta t = 0.1$~s is used to define displacements, although
the qualitative behavior is not too sensitive to this choice.
For low forcing, the non-Gaussian parameter is largest,
reflecting the extreme width of the blue data in
Fig.~\ref{distributions} as compared to the Gaussian fit.
$\alpha_2$ drops for larger $L$ in Fig.~\ref{non_gaussian}.  In
this respect the behavior is similar to glassy liquids, which are
more non-Gaussian close to the glass transition, and more Gaussian when in the liquid state \cite{kob97,debenedetti05,kob95a}.

\section{Conclusions}

In our experiments the granular bath particles become subdiffusive at
the lowest forcing $L$, while the floating particles are still
diffusive.  We conjecture this is because the floating particle
does not interact with the bath particles directly.  Typically
one thinks of subdiffusive behavior in particle systems as arriving due to crowding of
particles.  In our experiment, no matter how the bath
particles are crowded or interact with one another, the floating
particle sits atop them and does not experience the crowding.
It is interesting that the floating particle's long-time motion
continues being diffusive even though the particles in the bath
are not, and even though the floating particles' displacement
probability distributions are non-Gaussian.

As described above, the bath particles are forced periodically with
a period of 0.1~s, and this time scale appears to play a key role
in determining the floating particle motion.  It suggests that in
our experiment we have a novel source of random energy acting like
some sort of effective temperature.  This may provide insight into
other situations with non-thermal ``effective temperature'' baths.

\begin{acknowledgments}
Partial financial support by CONACyT, M\'exico, through grant 256176
(SEP-Ciencia B\'asica) is acknowledged.  The work of E.R.W. was
supported by the National Science Foundation (DMR-1609763).
\end{acknowledgments}

\end{document}